\documentstyle[mprocl]{article}

\def\be{\begin{equation}}
\def\ee{\end{equation}}
\def\bea{\begin{eqnarray}}
\def\eea{\end{eqnarray}}
\def\ket#1{\,|\, #1\negthinspace\negthinspace\rangle}
\def\bra#1{\langle\negthinspace\negthinspace #1\,|\,}
\def\vac{|{\rm vac}\rangle}

\begin{document}

\title{QUANTUM BLACK HOLES AS ATOMS}

\author{JACOB D. BEKENSTEIN }

\address{Racah Institute of Physics, The Hebrew University of
Jerusalem\\Givat Ram, Jerusalem, 91904 ISRAEL}

\maketitle\abstracts{
In some respects the black hole plays the same role in gravitation that 
the atom played in the nascent quantum mechanics.  This analogy suggests
that black hole mass $M$ might have a discrete spectrum.  I review the
physical arguments for the expectation that black hole horizon area
eigenvalues are uniformly spaced, or equivalently, that the spacing between
stationary black hole mass levels behaves like $1/M$.  This sort of spectrum
has also emerged in a variety of formal approaches to black hole quantization
by a number of workers (with some notable exceptions).  If true, this result
indicates a distortion of the semiclassical Hawking spectrum which could be
observable even for macroscopic black holes.  Black hole entropy suggests that
the mentioned mass levels should be degenerate to the tune of an exponential
in $M^2$, as first noted by Mukhanov.  This has implications for the
statistics of the radiation.  I also discuss open questions: whether radiative
decay will spread the levels beyond recognition, whether extremal black holes
can be described by this scheme, etc.  I then describe an elementary algebra
for the relevant black hole observables, an outcome of work by Mukhanov and
myself, which reproduces the uniformly spaced area spectrum.}
  
\section{Introduction}\label{sec:intro}

We have all been taught that at the Planck scale $M_{P}=
\hbar^{1/2} \approx 1.2 \times 10^{20}\, {\rm MeV}$ or
$\ell_{P}= \hbar^{1/2} \approx 1.6 \times 10^{-33}\,{\rm cm}$ (I use units
with Newton's $G=c=1$ and denote the charge of the electron by $-e$), quantum
gravity effects must become important.  But this scale is so extreme by
laboratory standards that it would seem one shall never be able to put quantum
gravity to the test in the laboratory.  One may, however, ask: is it possible
that by some recondite effect quantum gravity may make itself felt well below
the Planck energy (well above the Planck length) ?

Research in black hole physics has uncovered several mysteries: Why is the
statistical black hole entropy proportional to the horizon area ?  What
happens to the information in black hole evaporation ?  How to resolve those
mysteries in a simple way ?

Here I would like to discuss a scheme for eliciting some answers to
these questions. This is a partial and, as yet, informal scheme, not a quantum
theory of gravity.  It derives from an early attempt of mine~\cite{BekNC} to
quantize black holes, and a reformulation of it by Mukhanov.~\cite{Mukh86} In
classical general relativity the mass spectrum of black holes is a continuum.
The scheme suggests  that in quantum theory the black hole mass spectrum must
be discrete and highly degenerate in the sense that the black hole horizon
area is restricted to equispaced levels whose degeneracy corresponds, by the
usual Boltzmann--Einstein formula, to the black hole entropy associated with
each area eigenvalue.  The scheme makes use of rather common ideas from
quantum mechanics and field theory, while keeping the classical limit of
things in sight.~\cite{BekBrazil}  I shall begin with a physical discussion,
and become a bit formal in the sequel.

In setting out to give a quantum description of black holes, a primary
question (first asked by Wheeler in the late 1960's) is what is the complete
set of quantum numbers required to describe a black hole in a stationary
quantum state.  Quantum numbers are first and foremost attributes of
elementary particles.  Now an elementary object with mass below $M_P$ has its
gravitational radius tucked below its Compton wavelength; it is thus properly
termed ``elementary particle''.  By contrast an elementary object with mass
above $M_P$ has its Compton wavelength submerged under the gravitational
radius; it is best called a black hole.  The seeming discontinuity between the
two occasioned by the emergence of the horizon is not really there because at
the Planck scale the spacetime geometry should be quite fuzzy.  So there is no
in--between regime here, and by continuity the smallest black holes should be
quite like elementary particles, and should merit description by a few quantum
numbers like mass, charge, spin, {\it etc.\/}  

How do things change as the black hole gets larger ?  For macroscopic black
holes within general relativity, Wheeler~\cite{Wheeler} enunciated long ago 
his highly influential ``no hair'' conjecture:  a black hole is parametrized
only by quantities, such as mass, spin angular momentum and charge, which are
subject to Gauss--type laws.  One may add magnetic charge to Wheeler's list
because it is subject to a Gauss law, and there are generalization of the
Kerr--Newman solution with magnetic charge alongside electric charge.  I have
already argued~\cite{BekBrazil} that the many ``hairy'' black hole solutions
discovered in the last decade do not provide us with additional quantum
numbers for black holes.  Hence I assume here that the only quantum numbers of
a stationary black hole state are mass, electric charge, magnetic monopole and
spin.  

\section{Mass Spectrum of a Black Hole}\label{sec:mass_spectrum}

I shall thus focus on black hole eigenstates of mass $\hat M$, electric charge
$\hat Q$, magnetic monopole
$\hat G$, spin angular momentum $\hat {\bf J}^2$ and $\hat J_z$ and, of course,
linear momentum ${\hat {\bf P}}$. This last can be set to zero if one agrees to
work in the black hole's center of mass.  The eigenvalue spectra of $\hat Q,
\hat G,
\hat {\bf J}^2, {\hat J}_z$ are well known.  By making the standard assumption
that this last set of operators are mutually commuting, one may immediately
establish the mass spectrum for the extremal black holes.~\cite{Mazur}

The classical {\it extremal\/} Kerr--Newman black hole is defined by the 
constraint
\be
M^2=Q^2+G^2 +J^2/M^2
\label{eq:constraint}
\ee
Solving for $M$,  discarding the negative root solution (it gives
imaginary $M$), and replacing in this expression 
$Q\rightarrow qe$, $G\rightarrow g\hbar/2e$ and $J^2\rightarrow 
j(j+1)\hbar^2$ with $q, g$ integers and $j$ a nonnegative integer or
half--integer, one enforces the quantization of charge, magnetic monopole  and
angular momentum, and obtains the mass eigenvalues first found by
Mazur~\cite{Mazur}
\be
M_{qgj}= M_P\left[q^2 e^2/2 + g^2 \hbar^2/8 e^2 + \sqrt{(q^2 e^2/2 + g^2
\hbar^2/8 e^2)^2 + j(j+1)}\right]^{1/2}.
\label{eq:massextreme}
\ee

For nonextremal black holes I shall avail myself of the classical relation
between mass and area of the Kerr--Newman black hole:~\cite{ChrRuf}  
\be
M^2 = {A\over 16\pi}\left(1+{4\pi (Q^2 + G^2)\over A}\right)^2 + {4\pi
{\bf J}^2\over A}
\label{eq:CR}
\ee
One should note that only the parameter domain 
\be
A \ge \sqrt{(Q^2 + G^2)^2  + (8\pi)^2 {\bf J}^2}
\label{eq:restriction}
\ee
is physical because only there does the usual expression for $A$ as a function
of $M$, $Q$, $G$ and ${\bf J}$ hold. 

In converting Eq.~\ref{eq:CR} to a quantum relation between the operators
 $\hat M$, $\hat Q$, $\hat G$ and $\hat {\bf J}$ one faces the problem of
factor ordering.  Now the area of a black hole should be invariant under
rotations of its spin; since ${\bf \hat J}$ is the generator of such
rotations, one sees that $[\hat A, \hat{\bf J}] = 0$.  Similarly, area should
remain invariant under gauge transformation whose generator is, as usual, the
charge $\hat Q$.  Hence  $[\hat A, \hat Q] = 0$.  Duality invariance of the
Einstein--Maxwell equations would then suggest that $[\hat A, \hat G] =0$.
Hence one may merely replace the parameters in Eq.~\ref{eq:CR} by the
corresponding operators:
\be
\hat M^2 = \left[{\hat A\over 16\pi}\left(1+{4\pi (\hat Q^2 + \hat G^2)\over
\hat A}\right)^2 + {4\pi \hat {\bf J}^2\over \hat A}\right] \Theta\left (\hat
A-(\hat Q^2 + \hat G^2)^2  + (8\pi)^2 \hat {\bf J}^2\right)
\label{eq:spectrum}
\ee
This formula allows one to read off eigenvalues of $\hat M$ from those of 
$\hat A$, the charges and the angular momentum; it was first used in this
 sense long ago.~\cite{BekNC} The Heavyside $\Theta$ (step) function 
enforces the physical restriction Eq.~\ref{eq:restriction}; when this last is
violated, a zero mass eigenvalue is predicted, which means there is no such
black hole.

\section{Horizon Area as an Adiabatic Invariant}\label{sec:adiabatic}

Later in this presentation I derive the eigenvalues of $\hat A$
from a quantum operator algebra.  However, much insight can be
gleaned by using a simple physical approach inspired by the similarity of
horizon area to an adiabatic invariant in mechanics. What is an adiabatic
invariant ?

A physical system governed by a hamiltonian $H\left(q,p,\lambda\right)$ which
depends on a time dependent parameter $\lambda(t)$ is said to undergo an
adiabatic change if $\lambda$ varies on a timescale long compared to the 
longest timescale $T$ of the internal (oscillatory) motions : $\lambda^{-1}
d\lambda/dt << T^{-1}$.  Any dynamical quantity $A(q,p)$, a function or
functional of the $p$'s and $q$'s, which changes little during the time $T$
while $H$ accumulates a significant total change, is said to be an adiabatic
invariant.  Ehrenfest~\cite{Eh} showed that for a quasiperiodic system, all
Jacobi action integrals of the form $A\,=\,\oint p\, dq$ are adiabatic
invariants.  In particular,  for an harmonic oscillator with slowly
time--varying  frequency $\omega(t)$ (say a pendulum that oscillates with
small amplitude while the string is lengthened slowly), the Jacobi integral
equals
$2\pi E/\omega$.  Thus when the spring constant varies on a timescale $\gg
\omega^{-1}$, $E/\omega$ remains constant even while $E$ changes sizeably.

The subject is interesting here  because one can understand the
adiabatic invariance of $E/\omega$ in quantum terms.  For an harmonic
oscillator in a stationary state labeled by quantum number $n$, $E/\omega =
(n+{1\over 2})\hbar$.  One expects $n$ to remain constant during an adiabatic
change because the perturbations imposed on the system have frequencies $\ll
\omega$, so that  transitions between states of different $n$ are strongly
suppressed.  Therefore, the ratio $E/\omega$ is preserved. In the
Bohr--Sommerfeld theory (old quantum mechanics), all Jacobi actions are
quantized in integers: $\,\oint p\, dq=2\pi n\hbar$.  The above logic then
clarifies why the classical Jacobi actions are adiabatic invariants.

Ehrenfest generalized this insight:~\cite{Eh} any classical
adiabatic invariant (action integral or not) corresponds to a quantum entity
with discrete spectrum.  The rationale is that an adiabatic change, by virtue
of its slowness, is expected to lead only to continuous changes in the
system, not to jumps that change a discrete quantum number.  The preservation
of the value of the quantum entity would then explain the classical invariant
property.   

Ehrenfest's hypothesis can be used profitably in many problems. 
As an illustration consider a relativistic particle of rest mass $m$ and
charge $e$ spiralling in a magnetic field ${\bf B}$. One knows that the Larmor
spiralling frequency is
\be
\Omega={e|{\bf B}|\over \gamma_L m}={e|{\bf B}|\over E}
\label{eq:Larmor}
\ee
where $\gamma_L$ is Lorentz's gamma factor, and $E$ the total energy.  When
${\bf B}$ varies in space or in time slowly over one Larmor radius $r$ or
over one Larmor period $2\pi/\Omega$, there exists an adiabatic invariant
of the form 
\be
\Phi=\pi|{\bf B}|r^2,
\label{eq:flux}
\ee 
namely the magnetic flux through one loop
of orbit is invariant.~\cite{Jackson}  Now rewrite the energy
\be
E=m\left(1-\dot r^2-\dot z^2-r^2\Omega^2\right)^{-1/2}
\label{eq:energy}
\ee
by replacing $\dot z\rightarrow p_z/m\gamma_L$, taking into account that
$\dot r$ is nearly vanishing, and replacing $\Omega$ and $r$ by means of
Eq.~\ref{eq:Larmor} and
Eq.~\ref{eq:flux} to get
\be
 E^2=m^2+p_z^2+e^2 r^2 B^2 = m^2+p_z^2+e^2 \Phi B/\pi
\label{eq:temporary}
\ee
By Ehrenfest's hypothesis, in the quantum problem $\Phi$ should have a
discrete spectrum.  Thus  for fixed $p_z$, $E^2$ should be
quantized, possibly with uniformly spaced eigenvalues.  And indeed, the exact
solution of the relativistic Landau problem with the Klein--Gordon
equation~\cite{LL} leads to the spectrum
\be
E^2=m^2+p_z^2+e\hbar B (2n+1); \qquad n=0, 1, \cdots
\label{eq:Landaulevels}
\ee
which justifies the prediction from the Ehrenfest hypothesis.

I shall now argue that the horizon area of a nonextremal Kerr--Newman black
hole shows signs of being the analog of a mechanical adiabatic
invariant.~\cite{BHTrail}  Application of Ehrenfest's hypothesis in combination
with a dynamic calculation will then lead to the spectrum for the horizon area
operator.  

Consider a Reissner--Nordstr\"om black hole of mass $M$ and charge $Q$.   One
shoots in radially a classical point particle of charge $\varepsilon$ with
(conserved) total energy adjusted to the value $E=\varepsilon Q/r_{\cal H}$,
where $r_{\cal H}$ is the radius of the black hole in Boyer--Lindquist
coordinates.  In Newtonian terms this particle should marginally reach the
horizon where its potential energy just exhausts the total energy.   Study of
the exact equation of motion supports this conclusion: the particle's motion
has a turning point at the horizon.  Because of this, the assimilation of
particle by the black hole takes place especially slowly: it is an adiabatic
process.

Now the area of the horizon is originally 
\be
A = 4\pi {r_{\cal H}}^2  = 4\pi\left(M+\sqrt{M^2-Q^2}\right)^2
\label{eq:AreaRN}
\ee
and the (small) change inflicted on it by the absorption of the particle is
\be
\Delta A = {\theta_{RN}}^{-1} \cdot(\Delta M - Q\,\Delta Q/r_{\cal H}) 
\label{eq:dAreaRN}
\ee
with
\be
\theta_{RN} \equiv {1\over 2} A^{-1} \sqrt{M^2-Q^2} 
\label{eq:thetaRN}
\ee
Thus if the black hole is not extremal so that $\theta_{RN}\neq 0$, $\Delta
A=0$ because $\Delta M=E$ while $\Delta Q=\varepsilon$ and $E=\varepsilon
Q/r_{\cal H}$.  Therefore, the horizon area is invariant in the course of an
adiabatic change of the black hole.  But this conclusion does not extend to
the extremal black hole: when $Q=M$, $\sqrt{M^2-Q^2}$ in Eq.~\ref{eq:AreaRN}
is unchanged  to $O(\varepsilon^2)$ during the absorption, so that $\Delta A
=8\pi ME\neq 0$.

As a second example consider a Kerr black hole of mass $M$ and angular
momentum $J$. Send onto it a scalar wave of the form
$Y_{\ell\,m}(\theta,\phi)e^{-\imath\omega t}$.  It is known~\cite{Staro} that
for $\omega\approx \Omega m$, the absorption coefficient has the form
\be
\Gamma = K_{\omega\ell\,m}(M, J)\cdot \left(\omega -\Omega\, m\right)
\label{eq:gamma}
\ee
where
\be
\Omega\equiv {J/M\over {r_{\cal H}}^2  + (J/M)^2}
\label{eq:omega}
\ee
is the rotational angular frequency of the hole, while $K_{\omega\ell\,m} (M,
J)$ is a positive coefficient.  If one chooses $\omega=m\ \Omega$, the wave is
perfectly reflected.  By choosing $\ \omega-\Omega\, m\ $ slightly positive,
one arranges for a small fraction of the wave to get absorbed.  If the
reflected wave is repeatedly reflected back towards the black hole by  a
suitable large spheroidal mirror surrounding it, one can arrange for a sizable
fraction of the wave's energy and angular momentum to get absorbed
eventually.  But since this takes place over many cycles of reflection, the
change in the hole is an adiabatic one.

Now he horizon area of the Kerr black hole is
\be
A = 4\pi\left[\left(M+\sqrt{M^2-(J/M)^2}\right)^2+(J/M)^2\right]
\label{eq:AreaK}
\ee
and small changes of it are given by
\be
\Delta A = \theta_{K}{}^{-1}\cdot\left(\Delta M - \Omega \Delta J\right)
\label{eq:dAreaK}
\ee
where
\be
\theta_{K} \equiv {1\over 2} A^{-1} \sqrt{M^2-(J/M)^2} 
\label{eq:coeff}
\ee
The overall changes $\Delta M$ and $\Delta J$ must stand in the
ratio $\omega/m$.  This can be worked out from the energy--momentum tensor,
but is simplest appreciated by thinking of the wave as made of quanta, each 
with energy $\hbar\omega$ and angular momentum $\hbar m$.  Since I chose
$\omega\approx \Omega\, m$, it follows from Eq.~\ref{eq:dAreaK} that if the
black hole is not extremal, $\Delta A \approx 0$ to the accuracy of the former
equality.  Evidently, here too horizon area is invariant during adiabatic
changes.  This conclusion is inapplicable to the extremal black hole for
reasons similar to those in our first example. 

The two examples, and the one to follow in the next section, support the
conjecture that for a nonextremal Kerr-Newman black hole, the horizon area $A$
is, classically, an adiabatic invariant.   By taking Ehrenfest's hypothesis 
seriously, I conclude that the horizon area of a nonextremal Kerr-Newman
quantum  black hole, $\hat A$, just like that of the extremal black hole,
must have a discrete eigenvalue spectrum.

\section{Spacing and Multiplicity of the Area Levels}\label{sec:spacing}

In view of the last conclusion I write the horizon area eigenvalues as
\be
a_n = f(n);\qquad n=1, 2, 3, \cdots
\label{eq:eigenareas}
\ee
The function $f$ must clearly be positive and monotonically increasing (this
last just reflects the ordering of eigenvalues by magnitude).  However, one
cannot infer from all this that $f$ is linear in its argument. For not  every
adiabatic invariant has a simply quantized spectrum: if K is an adiabatic
invariant, and its quantum counterpart $\hat K$ has a uniformly spaced
spectrum, $K^2$ is still an adiabatic invariant, but $\hat K^2$'s spectrum is
not uniformly spaced.  At any rate, in light of Eq.~\ref{eq:CR} and the
quantization of charge, magnetic monopole, and angular momentum, this result
implies that the nonextremal Kerr-Newman black hole mass has a discrete
spectrum.  Its form will be elucidated in Sec.~\ref{sec:demystify}.

To elucidate the spacing of the area levels, I elaborate here on
Christodoulou's prescient question:~\cite{ChrRuf}  can the assimilation of a
point particle by a Kerr black hole be made reversibly in the sense that all
changes of the black hole are undone by absorption of a suitable second
particle ?  This is a good question because black hole horizon area
cannot decrease,~\cite{HawkingArea} so that any process which causes it to
increase is irrevocably irreversible.  Christodoulou's answer, later
generalized to the Kerr--Newman black hole,~\cite{ChrRuf} was that the
process is reversible if the (point) particle, which may be electrically
charged and carry angular momentum,  is injected at the horizon from a
turning point in its orbit.  In this case the horizon area (or equivalently
the irreducible mass) is left unchanged, so that the effects on the black
hole can be undone by a second reversible process which adds charges and
angular momentum opposite in sign to those added by the first.  Our
Reissner--Nordstr\"om example in Sec.~\ref{sec:adiabatic} is a special case of a
Christodoulou reversible process.  For Kerr--Newman black holes,
Christodoulou's reversible process furnishes one with a further example of the
adiabatic invariance of horizon area (after all, the capture from a turning
point is slow - adiabatic).  One can check that Christodoulou
and Ruffini's calculation proves reversibility only for nonextremal black
holes.  

All the preceding is purely classical theory: in the Christodoulou--Ruffini
process the particle follows a classical path, and must be a point particle
in order for its absorption to leave the area unchanged.  How would quantum
theory modify things ?  I do not intend anything so complicated as
solution of a Schr\"odinger--like equation.  But as a concession to quantum
theory let me ascribe to the particle a finite radius $b$ while continuing to
assume, in the spirit of Ehrenfest's theorem in quantum mechanics, that the
center of mass of the particle follows a classical trajectory.  Recalculating
{\it a la\/} Christodoulou--Ruffini for a particle of mass $\mu$ one finds
that absorption of the particle now necessarily involves an increase in area. 
This is minimized if the particle is captured when its center of mass is at a
turning point of its motion a proper distance $b$ away from the
horizon:~\cite{BekEntropy}
\be
(\Delta A)_{\rm min} = 8\pi \mu b.
\label{eq:minimum}
\ee

This last conclusion  fails for {\it extremal\/} black holes because the
analog of the quantity $\theta_K$ in Eq.~\ref{eq:coeff} diverges in that case.
The minimal increase in area is not Eq.~\ref{eq:minimum}, but a quantity
dependent on $M$, $Q$, $G$ and $J$. 

The limit $b\rightarrow 0$ of Eq.~\ref{eq:minimum} recovers Christodoulou's
reversible process for the nonextremal black holes.  However, a quantum point
particle is subject to quantum uncertainty.  If it is known to be at the
horizon with high accuracy, its radial momentum is highly uncertain; this
prevents the turning point condition from being fulfilled.  And, of course, a
relativistic quantum point particle cannot even be localized to better than a
Compton length
$\hbar/\mu$.  Thus in quantum theory the limit $b\rightarrow 0$ is not a
legal one. One can get an idea of the smallest possible increase in horizon
area in the quantum theory by putting $b\rightarrow \xi\hbar/\mu$ in
Eq.~\ref{eq:minimum}, where $\xi$ is a number of order unity.  Thus
\be
 (\Delta A)_{\rm min} =  8\pi\xi \hbar = 
\alpha{\ell_{P}}^2
\label{eq:universal}
\ee
Surprisingly, for nonextremal black holes, $(\Delta A)_{\rm min}$ turns out
to be independent of the black hole characteristics $M, Q, G$ and $J$. 
   
The fact that for nonextremal black holes there is a minimum area increase
as soon as one allows quantum nuances to the problem, suggests that this 
$(\Delta A)_{\rm min}$ corresponds to the spacing between eigenvalues of 
$\hat A$ in the quantum theory.  And the fact
that $(\Delta A)_{\rm min}$ is a universal constant suggests that the spacing
between eigenvalues is a uniform spacing.  For it would be strange indeed if
that spacing were to vary, say, as mass of the black hole, and yet the
increment in area resulting from the best approximation to a reversible
process would contrive to come out universal, as in Eq.~\ref{eq:universal}, by
involving a number of quantum steps inversely proportional to the eigenvalue
spacing.  I thus conclude that for nonextremal black holes the spectrum of
$\hat A$ is
\be
a_n = \alpha\, {\ell_{P}}^2\, (n+\eta);\quad\ \eta>-1; \quad n = 1,
2,\cdots
\label{eq:areaspectrum} 
\ee
where the condition on $\eta$ excludes nonpositive area eigenvalues. 

Now Eqs.~\ref{eq:universal}-\ref{eq:minimum} fail for an extremal Kerr--Newman
black hole, so one cannot deduce as above that its area eigenvalues are evenly
spaced.  This is entirely consistent with Eq.~\ref{eq:massextreme} according
to which the area spectrum is then very complicated. 

\section{Demystifying Black Hole's Entropy Proportionality to
Area}\label{sec:demystify}

The reader will have noticed that the previous arguments have
said nothing about entropy; the discussion has been at the level of
mechanics, not statistical physics. I shall now make use of
Eq.~\ref{eq:areaspectrum} to understand, in a pleasant and intuitive way, the
mysterious proportionality between black hole entropy and horizon area. 

The quantization of horizon area in {\it equal\/} steps brings to mind
an horizon formed by patches of equal area $\alpha\ell_P^2$ which get added
one at a time.  It is unnecessary - even detrimental - to think of a specific
shape or localization of these patches.  It is their standard size which is
important, and which makes them all equivalent.   This patchwork horizon can
be regarded as having many degrees of freedom, one for each patch.  After
all, the concept ``degree of freedom'' emerges for systems whose parts can act
independently, and here the patches can be added to the patchwork one at a
time.  In quantum theory degrees of freedom independently manifest distinct
states.  Since the patches are all equivalent, each will have the same number
of quantum states, say, $k$.  Therefore, the total number of quantum states
of the horizon is
\be
N = k^{A/\alpha \ell_P^2}
\label{eq:numberstates}
\ee
where $k$ is a positive  integer and the effects of the $\eta$ zero point in
Eq.~\ref{eq:areaspectrum} are glossed over in this, intuitive, argument.

Nobody assures one that the $N$ states are all equally probable.  But if I
assume that the $k$ states of each patch--degree of freedom are all equally
likely, then all $N$ states are equally probable.  In that case the statistical
(Boltzmann) entropy associated with the horizon is $\ln N$ or
\be
S_{BH} = {\ln k\over \alpha}{A\over \ell_P^2}
\label{eq:BHentropy}
\ee
Thus is the proportionality between black hole entropy and horizon
area justified in simple terms. One may interject that even if not all $k$
states are equally probable, one can still use Eq.~\ref{eq:BHentropy}
provided $k$ is regarded as an effective number of equally probable states.
Comparison of Eq.~\ref{eq:BHentropy} with Hawking's coefficient in the black
hole entropy calibrates the constant $\alpha$: 
\be
\alpha=4\ln k
\label{eq:calibration}
\ee

The above argument is meant to demystify the direct proportionality of
black hole entropy and horizon area.  It depends crucially on the uniformly
spaced area spectrum.  The logic leading to the number of states
Eq.~\ref{eq:numberstates} has occasionally been used without regard to any
particular area spectrum,~\cite{BekinWheeler,Sorkin} but these early arguments
lack conviction because their partition of the horizon into equal area cells
would be without basis if the desired result (entropy $\propto$ area) were not
known.

Mukhanov's~\cite{Mukh86,BekMukh} alternate route to 
Eqs.~\ref{eq:numberstates} and \ref{eq:calibration} starts from the accepted
formula relating black hole area and entropy. In the spirit of the
Boltzmann--Einstein formula, he views
$\exp(S_{BH})$ as the degeneracy of the particular area  eigenvalue because
$\exp(S_{BH})$ quantifies the number of microstates of the black hole
that correspond to a particular external macrostate.  Since black hole entropy
is determined by thermodynamic arguments only up to an additive constant, one
writes, in this approach,
$S_{BH}= A/4{\ell_{P}}^{2}+$ const. Substitution of the area eigenvalues 
from Eq.~\ref{eq:areaspectrum} gives the degeneracy corresponding to
the $n$--th area eigenvalue:
\be
g_n = \exp\left({a_n\over 4{\ell_{P}}^2} + {\rm const.}\right) = 
g_1\ e^{\alpha
(n-1)/4}
\label{eq:exp}
\ee

As stressed by Mukhanov, since $g_n$ has to be integer for every $n$,
this is only possible when~\cite{BekMukh,QG}
\be
g_1 = 1, 2, \cdots\quad{\rm and}\quad \alpha = 4\times\left\{\ln 2,\ln 3, 
\cdots\right\} 
\label{eq:options}
\ee
The simplest option for $g_1$ would seem to be $g_1 = 1$ (nondegenerate
black hole ground state).  Here the additive constant in
Eq.~\ref{eq:exp} must be retained: were it zero, the area $a_1$ would also
vanish which seems an odd thing for a black hole.  Just this case was studied
in Ref.~16; it is a bit ugly in that the eigenvalue law
Eq.~\ref{eq:areaspectrum} and the black hole entropy include related but
undetermined additive constants.  

The next case, $g_1=2$ (doubly degenerate black hole ground state), no
longer requires the ugly additive constant in the black hole entropy to keep
$a_1$ from vanishing.  With this constant set to zero and the choice
$\alpha=4\ln 2$ corresponding to $k=2$, Eqs.~\ref{eq:areaspectrum}  and
\ref{eq:exp} require that  $\eta=0$ so that one is rid of the second ugly
constant as well.  The area spectrum is
\be
a_n = 4{\ell_{P}}^2\ln 2\cdot n;   \quad  n=1,2,\cdots
\label{eq:truespectrum}
\ee
This spectrum, which I shall adopt henceforth, is good for nonextremal
Kerr--Newman black holes.   The corresponding degeneracy of area eigenvalues
\be
g_n = 2^n
\label{eq:degeneracy}
\ee
corresponds to a doubling of the degeneracy as one passes from one area
eigenvalue to the next largest.  Mukhanov~\cite{Mukh86} thought of this
multiplicity as the number of ways in which a black hole in the $n$--th area
level can be made by first making a black hole in the ground state,
and then proceeding to ``excite it'' up the ladder of area levels in
all possible ways.  Danielsson and Schiffer~\cite{DanSch} considered
this multiplicity as representing rather the number of ways the black hole
with area $a_n$ can ``decay'' down the staircase of levels to the ground state. 

To what extent do these intuitively physical predictions correspond to formal
results from existing quantum gravity schemes ?  The uniformly spaced area
spectrum was first proposed in 1975.~\cite{BekNC}  Starting with Kogan's 1986
string theoretic argument,~\cite{Kogan} a number of formal calculations, most
in the last few years, have recovered this form of the spectrum.  Mention may
be made of the quantum membrane approaches of Maggiore~\cite{Maggiore}
and of Lousto~\cite{Lousto} which establish the uniformly spaced levels as the
base for excitations of the black hole.  

There are also several canonical quantum gravity treatments of a shell or  ball
of dust collapsing on its way to black hole formation.  Those by
Schiffer~\cite{Schiffer} and Peleg~\cite{Peleg} get the uniformly spaced area
spectrum.  But Berezin,~\cite{Berezin} as well as Dolgov and
Khriplovich,~\cite{DolgKhrip} obtain mass spectra for the ensuing black hole
which correspond to discrete area spectra with {\it nonuniform\/} spacing (in
Berezin's approach the levels are infinitely degenerate).  Other  canonical
quantum gravity approaches by  Louko and M\"akel\"a,~\cite{LoukMak}
Barvinskii and Kunstatter,~\cite{BarvKunst} M\"akel\"a~\cite{Makela} and
Kastrup~\cite{Kastrup} treat a spherically symmetric vacuum spacetime
endowed with dynamics by some subtlety, and also come up with a uniformly
spaced area spectrum.  There is, however, no general agreement on the spacing
of the levels.  The analogous treatment of the charged black hole by
 M\"akel\"a and Repo~\cite{MakelaRepo} gets a nonuniform area spectrum.     

Mention must also be made of the loop quantum gravity determination of the
black hole area spectrum by Barreira, Carfora and Rovelli~\cite{Rovelli} and
by Krasnov.~\cite{Krasnov}  It leads to a discrete spectrum of complex form
and highly {\it nonuniform\/} spacing.  An exception to this qualification is
the spectrum for the extremal neutral Kerr black hole where the
aforementioned determination, assuming it can be applied to a nonspherical
black hole, would concur with the Mazur spectrum Eq.~\ref{eq:massextreme} (I.
Khriplovich pointed this out to me during the meeting).

This set of contradictory conclusions seems to certify the view held by many
that none of the existing formal schemes of quantum gravity is as yet a
quantum theory of gravity.

\section{The Black Hole Line Emission Spectrum}\label{sec:lines}

The particular area spectrum Eq.~\ref{eq:truespectrum} implies, by virtue of
the Christodoulou--Ruffini relation Eq.~\ref{eq:CR}, a definite discrete mass
spectrum.  For zero charge and angular momentum the mass spectrum is of the
form
\be
M\propto \surd n; \qquad n=1, 2
\label{eq:massspectrum}
\ee
implying the level spacing
\be
\omega_0\equiv \Delta M/\hbar = ( 8\pi M)^{-1}\ln 2
\label{eq:Bohr}
\ee 
This simple result is in agreement with Bohr's correspondence principle:
``transition frequencies at large quantum numbers should equal classical
oscillation frequencies'', because a classical Schwarzschild black hole
displays `ringing frequencies' which scale as $M^{-1}$, just as
Eq.~\ref{eq:Bohr} would predict.  This agreement would be destroyed if the area
eigenvalues were unevenly spaced.  Indeed, the loop gravity spectrum mentioned
in Sec.~\ref{sec:demystify} fails this correspondence principle test.

By analogy with atomic transitions, a black hole at some particular mass
level would be expected to make a transition to some lower level with
emission of a quantum (or quanta) of any of the fields in nature.  The
corresponding line spectrum - very different from the Hawking semiclassical
continuum - was first discussed in Ref.~1 and further analyzed much
later.~\cite{BekMukh,QG}   It comprises lines with all frequencies which are
integral multiples of $\omega_0$.  An elementary estimate gives the strength
of the successive lines as falling off roughly as
$\exp(-8\pi M\omega/\hbar)$, so that only a few lines - those within the
Hawking peak of the semiclassical emission - will be easily visible.   The
statistics of quanta in the radiation are reasonable: the number of quanta of
a given kind in a given line  emitted over a fixed time interval is Poisson
distributed.

Most important, as Mukhanov was first to remark,~\cite{BekMukh,QG}
this simple spectrum provides a way to make quantum gravity effects
detectable even for black holes well above the Planck mass:  the uniform
frequency spacing of the black hole lines occurs at all mass scales, and the
unit of spacing is inversely proportional to the black hole mass over all
scales.  Of course, for very massive black holes, one would expect all the
lines to become dim and  unobservable (just as in the semiclassical
description the Hawking radiance intensity goes down as
$1/M^2$), but there should be a mass regime (primordial mini--black holes ?)
well above Planck's for which the first few uniformly spaced lines should be
detectable under optimum circumstances.  It is thus clearly important to
understand clearly the nature of the line spectrum.

The first natural question is whether natural broadening of the lines
will not smear the spectrum into a continuum. First explored by
Mukhanov,~\cite{Mukh86} this issue has been revisited recently by both of
us.~\cite{BekMukh,QG}  By the usual argument the broadening of
a line,
$\delta\omega$, should be of order
$\tau$, the typical time (as measured at infinity) between transitions of the
black hole from level to level.  One may thus estimate the rate of loss of
black hole mass as
\be
{dM\over dt} \approx -{\hbar\omega_0\over \tau}=- {\hbar\,\ln 2\over 8\pi M\,
\tau }
\label{eq:rate}
\ee
Alternatively, one can estimate $dM/dt$ by assuming, in accordance with 
Hawking's  semiclassical result, that the radiation is black body radiation,
at least in its intensity.  Taking the radiating area as 
$4\pi (2M)^2$ and the temperature as  $\hbar/8\pi M$ one gets
\be
{dM\over dt} = - {\gamma\hbar\over 15360\pi M^2}
\label{eq:blackbody}
\ee
where $\gamma$ is a fudge factor that summarizes the grossness of our
approximation.  By comparing Eq.~\ref{eq:blackbody} with Eq.~\ref{eq:rate} one
infers $\tau$ which then gives
\be
{\Delta\omega\over \omega_o} \sim 0.019\,\gamma
\label{eq:relative}
\ee

Mukhanov and I regard $\gamma$ to be of order unity, which would make
the natural broadening weak and the line spectrum sharp.  More recently
M\"akel\"a~\cite{Makela2} has estimated a much larger value, and claimed
that the line  spectrum effectively washes out into a continuum.  He
views this as a welcome development because it brings the ideas about black
hole quantization, as here described, into consonance with Hawking's
smooth semiclassical spectrum.  

M\"akel\"a uses Page's~\cite{Page} estimate of black hole luminosity which
takes into account the emission of several species of quanta, whereas our
value $\gamma \sim 1$ is based on one species.  It is, of course, true that a
black hole will radiate all possible species, not just one.  This is expected
to enhance $\gamma$ by an order or two over the naive value.   But it is also
true that because the emission is, in the first instance, in lines, part of the
frequency spectrum is thus blocked, which should lead to a reduced value
for $\gamma$ in Eq.~\ref{eq:blackbody}.  Mukhanov and I consider the two
tendencies to partly compensate, and expect $\gamma$ to exceed its putative
value of unity by no more than an order of magnitude.  According to
Eq.~\ref{eq:relative} this should leave the emission lines unblended.

The above is not to say that the emission spectrum should be purely a
line spectrum. Multiple quanta emission in one transition will also contribute
a continuum.  To go back to atomic analogies, the transition from the 2s to
the 1s states of atomic hydrogen, being absolutely forbidden by one--photon
emission, occurs with the long lifetime of 8 s by two--photon emission; the
photon spectrum is thus a continuum over the relevant frequency range. 
Likewise, some multiple quanta emission should accompany  transitions of the
black hole between its mass levels thus forming a continuum that would compete
with the line spectrum~\cite{BekMukh,QG} (at the conference G. Lavrelashvili
reminded me of this).  However, no reason is known why one--quantum transition
would be forbidden in the black hole case.  Thus my expectation, again based
on the atomic analogy, is that most of the energy will get radiated in
one--quantum transitions which give lines.  Thus the spectrum, in first
approximation, should be made up of lines sticking quite clearly out of a
lowly continuum.

In atomic physics emission spectra display a hierarchy of splittings which
can be viewed as reflecting the hierarchical breaking of the various
symmetries.  Thus in atomic hydrogen the $O(4)$ symmetry of the Coulomb
problem, which is reflected in the Rydberg--Bohr spectrum, is broken by
relativistic effects (spin--orbit interaction and Thomas precession) thus
giving rise to fine structure splitting of lines.  But even an exact
relativistic treatment in the framework of Dirac's equation leaves the 2s and
the 2p levels perfectly degenerate.  They are split by a minute energy by
vacuum polarization effects and the Lamb shift. In addition, the rather
weak interaction of the electron with the nuclear proton's magnetic moment
leads to a small hyperfine splitting of members of some of the other fine
structure multiplets.  The very simple spectrum Eq.~\ref{eq:massspectrum} is
analogous to the hydrogenic Rydberg--Bohr spectrum.  Are there any splittings
of the lines here discussed ?

There is certainly room for splitting because of the $2^n$--fold degeneracy 
of the levels.  The question is whether there is some breaking of
symmetry, analogous to the ones taking place in the atomic case, which would
actually split the black hole lines.  To answer the question one obviously
needs more detailed information about the way the black hole mass spectrum
arises in the context of the various symmetries than we have heretofore
elicited from our simple arguments.  The desire to find out more about this
question is the main impetus behind the algebraic approach to be described in
Sec.~\ref{sec:algebraic}

Before leaving the subject I mention some related puzzles which cannot
be fully resolved purely by analogy with atomic physics.  For instance,
according to Eq.~\ref{eq:Bohr}, a Schwarzschild black hole cannot emit quanta 
at frequencies below $\omega_0$; by the usual argument of microscopic
reversibility, it should not absorb below this frequency.  However,
classically a Schwarzschild black hole absorbs all frequencies, albeit with
decreasing crossection as the frequencies become small compared to
$M^{-1}$.  Ref.~1 offered a solution to this paradox based on the
observation that at the low frequencies envisaged, the {\it classical\/}
absorptivity of the hole is so small that the expected amount of energy
absorbed is always below one quantum's worth, unless the energy of
the incident wave much exceeds $\hbar\omega_0$.  Thus the classical
description, which conflicts with the quantum description, must fail
unless enough energy is incident to elicit a quantum jump of the black hole
by {\it many quanta\/} absorption (analogous to many--photon processes in
nonlinear optical media).  This anomalous absorption would be interpreted in
classical theory as the expected absorption of sub--treshold frequencies.

The above resolution, by invoking a many--quantum process, brings back
the specter of many quantum emission and an emission continuum to compete
with the line spectrum, and makes the more urgent the task of proceeding
beyond mere analogies in the description of the emission process.

\section{Algebraic Approach to the Quantum Black Hole}\label{sec:algebraic}

In quantum theory one usually obtains spectra of operators from the algebra 
they obey.  For instance, Pauli~\cite{Pauli} obtained the complete
spectrum of hydrogen in nonrelativistic theory from the $O(4)$ algebra
of the relevant operators.  This approach sidesteps the question of
constructing the wavefunctions for the states.  I will now describe an
axiomatic algebra, whose genesis goes back to joint work with Mukhanov, which
describes the quantum black hole and gives an area spectrum identical to the
one found above. It thus supports the results obtained previously, and
illuminates the question of level splitting.~\cite{BekBrazil}  

Sec.~\ref{sec:mass_spectrum} introduced some of the relevant operators for a
black hole: mass $\hat M$, horizon area $\hat A$, charge $\hat Q$, monopole
$\hat G$ and angular momentum $\hat {\bf J}$.  The spectrum of $\hat Q$ is
$\{qe;\ q={\rm integer}\}$, that of $\hat {\bf J}^2\ \{j(j+1)\hbar^2\}$, while
that of $\hat J_z$ is $ \{-j\hbar,\ -(j-1)\hbar,\ \cdots,\ (j-1)\hbar,\ 
j\hbar\}$ with $j$ a nonnegative integer or half--integer.  In all that
follows I shall ignore $\hat G$ for brevity.  The first
axiom is:

\medskip
\noindent {\bf Axiom 1}: Horizon area is represented by a {\it positive 
semi-definite\/} operator $\hat A$ with a {\it discrete\/} spectrum $\{a_n;
\ n=0, 1, 2\cdots\ \}$.  The degeneracy of the eigenvalue $a_n$, denoted
$g(n)$, is independent of the $j, m$ and $q$.
\medskip

Discreteness of the area spectrum, as suggested by the adiabatic invariant
character of horizon area, is formalized in this axiom; it is not here
proved.  One imagines the eigenvalues to be arranged so that $a_0=0$
corresponds to the vacuum $\vac$ (state devoid of any black holes) 
while the rest of the $a_n$ are arranged in  order of increasing value.
Since I do not refer to $\hat G$ in what follows, no confusion will arise
with the use of $g$ for degeneracy. I take $g(0)=1$.

Because $\hat A$, $\hat Q$,  ${\bf \hat J}^2$ and  $\hat J_z$ mutually
commute, one can infer the spectrum of $\hat M$ from that of $\hat A$ 
directly from the Christodoulou--Ruffini formula Eq.~\ref{eq:CR} which, 
as mentioned in Sec.~\ref{sec:mass_spectrum}, does not suffer from factor
ordering ambiguities. But the triviality of the algebra precludes one learning
anything about the spectrum of $\hat A$ itself.

This motivates the introduction of creation operators for black
holes in their various states.  In view of the similarities between black hole
and elementary particles, it seems not farfetched to imagine black holes as
particles of some field, and then creation operators appear naturally.
Our second axiom is thus
\medskip

\noindent {\bf Axiom 2}: There exist operators $\hat R_{njmqs}$  with the
property that  $\hat R_{njmqs}\vac$ is a {\it one\/}
black hole state with horizon area $a_n$, squared spin
$j(j+1)\,\hbar^2$, $z$--component of spin $m\,\hbar$, charge $qe$ and
internal quantum number $s$.  All one--black hole states are spanned by the
basis $\{\hat R_{njmqs}\vac\}$.
\medskip

I do not purport to construct $\hat R_{njmqs}$.  The internal quantum numbers
are necessary because from the black hole entropy one knows that each state
seen by an external observer corresponds to many internal states; these need
to be distinguished by additional quantum numbers (below called variously $s,
t$ or $r$). In the interest of clarity in the equations, I shall, when no
misunderstanding can arise, write $\hat R_{\kappa\, s}$ or just  $\hat
R_{\kappa}$ for $\hat R_{njmqs}$.

Commutation of the operators now available creates more operators.  If this
process continues indefinitely, no information can be obtained from the
algebra unless additional assumptions are made.  Faith that it is possible to
elucidate the physics from the algebra leads me to require closure of the
algebra (which I suppose to be linear in analogy with many physically
successful algebras) at an early stage.  This is formalized in

\medskip
\noindent {\bf Axiom 3}: The operators $\hat A,\ \hat {\bf J},\ \hat Q,\ \hat
R_{\kappa s}$ and $[\hat A,\hat R_{\kappa s}]$ form a closed, linear,
infinite--dimensional nonabelian algebra.
\medskip

Now the physical interpretation of the operators $\hat R_{\kappa\, s}$ leaves
little choice regarding their commutators with the operators $\hat A$, $\hat
Q$ and  ${\bf \hat J}$.  The algebra of $\hat A$ will be the subject of the
Sec.~\ref{sec:area_algebra}.  Here I start with  ${\bf \hat J}$.  Since $\hat
R_{njmqs}\vac$ is defined as a state with spin
quantum numbers $j$ and $m$, the collection of such states with fixed $j$
and all allowed $m$ must transform among themselves under rotations of the
black hole like the spherical harmonics $Y_{jm}$ (or the corresponding
spinorial harmonic when $j$ is half--integer).  Since  $\vac$ must obviously
be invariant under rotation, one learns that the $\hat R_{njmqs}$ may be taken
to behave like an irreducible spherical tensor operator of rank $j$ with the
usual $2j+1$ components labeled by $m$.~\cite{Merz}  This means that
\be
[\hat J_z, \hat R_{\kappa}] = m_\kappa\,\hbar\, \hat R_{\kappa}
\label{eq:commuteJz}
\ee
and
\be
[\hat J_\pm, \hat R_{\kappa}] =
\sqrt{j_\kappa(j_\kappa+1)-m_\kappa(m_\kappa\pm 1)}\,\hbar\,\hat R_{\kappa\,
m_\kappa\pm 1}
\label{eq:commuteJ+}
\ee
where $\hat J_\pm$ are the well known raising and lowering operators for
the $z$-component of spin.  To check these commutators I first operate with
Eq.~\ref{eq:commuteJz} on  $\vac$ and take into account that $\hat {\bf J}
\vac\ = 0$ (the vacuum has zero spin) to get
\be
\hat J_z\,\hat R_{\kappa s}\vac = m_\kappa\hbar \hat
R_{\kappa s}\vac
\label{eq:eigenvalueJz}
\ee
Also from the relation~\cite{Merz} $\hat {\bf J}^2 = (\hat J_+\hat J_- +
\hat J_-\hat J_+)/2 +\hat J_z^2$, one can work out $[\hat {\bf
J}^2, \hat R_{\kappa s}]$ and operate with it on $\vac$;
after double use of Eqs.~\ref{eq:commuteJz} and \ref{eq:commuteJ+} one gets
\be
\hat {\bf J}^2\,\hat R_{\kappa s}\vac\, = j_\kappa(j_\kappa+1)\hbar^2 \hat
R_{\kappa s}\vac
\label{eq:eigenvalueJ^2}
\ee
Of course both of these results were required by the definition of 
$\hat R_{njmqs}\vac$.

Moving on one recalls that $\hat Q$ is the generator of (global) gauge
transformations of the black hole, which means that for an arbitrary real
number $\chi$, $\exp(\imath\chi\hat Q)$ elicits a phase change $\chi$ of the
black hole state:
\be
\exp(\imath\chi \hat Q)\, \hat R_{\kappa
s}\vac\, = \exp(\imath\chi q_\kappa e)\,\hat R_{\kappa s}\vac:
\label{eq:gauge}
\ee
 This equations parallels
\be
\exp(\imath\phi \hat J_z/\hbar)\, \hat R_{\kappa
s}\vac\, = \exp(\imath\phi m_\kappa)\,\hat R_{\kappa
s}\vac,
\label{eq:rotation}
\ee
which expresses the fact that $\hat J_z$ is the generator of rotations of
the spin about the $z$ axis.  Thus by analogy with Eq.~\ref{eq:commuteJz} 
one may settle on the commutation relation
\be
[\hat Q, \hat R_{\kappa s}] = q_\kappa e R_{\kappa s}.
\label{eq:commuteQ}
\ee
Operating with this on the vacuum (recall that $\hat Q\vac = 0$) gives
\be
\hat Q\, \hat R_{\kappa s}\vac = q_\kappa e
R_{\kappa s} \vac
\label{eq:charge}
\ee
so that $R_{\kappa s}\vac$ is indeed a one black hole state with definite
charge $q_\kappa e$, as required.

In addition to Eqs.~\ref{eq:commuteJz}-\ref{eq:commuteJ+} and
\ref{eq:commuteQ} one would like to determine $[\hat A,
R_{\kappa s}]$, but since it is unclear what kind of symmetry transformation
$\hat A$ generates, a roundabout route is indicated.

\section{Algebra of the Area Observable}\label{sec:area_algebra}

Consider the Jacobi identity
\be
[\hat A, [\hat V, \hat R_{\kappa}]] + [\hat V, [\hat R_{\kappa},
\hat A]] + [\hat R_{\kappa}, [\hat A, \hat V]]=0
\label{eq:Jacobi}
\ee
valid for three {\it arbitrary\/} operators $\hat A$, $\hat V$ and $\hat
R_{\kappa s}$.  Suppose one replaces $\hat V$ in turn by $\hat J_z$, 
$\hat J_\pm$ and $\hat Q$, and makes use of 
Eqs.~\ref{eq:commuteJz}-\ref{eq:commuteJ+} and \ref{eq:commuteQ} as well as
the commutativity of $\hat J_z, \hat J_\pm,$ $\hat Q,$ and $\hat A$ to obtain
the three commutators
\bea
& & [\hat J_z, [\hat A, \hat R_{\kappa\, m_\kappa}]]  =  m_\kappa\,\hbar\,
[\hat A, R_{\kappa\, m_\kappa}], \nonumber \\
 & & [\hat J_\pm, [\hat A, \hat R_{\kappa\, m_\kappa}]]  = 
\sqrt{j_\kappa(j_\kappa+1)-m_\kappa(m_\kappa\pm 1)}\,\hbar\,[\hat A,\hat
R_{\kappa\, m_\kappa\pm 1}], \nonumber \\
 & & [\hat Q, [\hat A, \hat R_{\kappa\, m_\kappa}]]  =  q_\kappa e\, [\hat A,
R_{\kappa\, m_\kappa}].
\label{eq:three}
\eea
Thus, for fixed $j, m,$ and $q$, a particular $[\hat A, \hat R_{njmqs}]$  has
commutators of the same form as all the $\hat R_{njmqs}$ with various $n$ and
$s$.  Hence one can write generically
\be 
[\hat A, \hat R_{\kappa s}]=\sum_{n_\lambda t} h_{\kappa s}{}^{\lambda t}\,
\hat R_{\lambda t} + \hat T_{\kappa s}
\label{eq:commuteA} 
\ee
where $n_\lambda$ belongs to the set $\lambda$, the $h_{\kappa\,
s}{}^{\lambda\, t}$ are constants and $\hat T_{\kappa s}$ is an operator
independent of all the $\hat R_{\kappa s}$ (for otherwise it could be lumped
with them in the r.h.s.).   Eq.~\ref{eq:commuteA} is really a definition of
$\hat T_{\kappa s}$, which operator obviously mimics the behavior of $\hat
R_{\kappa s}$ under rotations and gauge transformations.

Operating with Eq.~\ref{eq:commuteA} on the vacuum (and remembering that $\hat
A\vac\, = 0$ because of the postulated $a_0 =0$) one gets
\be
a_\kappa\,\hat R_{\kappa s}\vac\, = \sum_{n_\lambda t} h_{\kappa
s}{}^{\lambda t}\, \hat R_{\lambda t}\vac +
\,\hat T_{\kappa s}\vac
\ee
Now because the $\hat R_{\lambda t}\vac$ with various $n_\lambda$ and $t$ are
independent, one must set
\be
h_{\kappa s}{}^{\lambda t}\,  =a_\kappa\,
\delta_{n_\kappa}{}^{n_\lambda}\,\delta_s{}^t \qquad {\rm and}\qquad \hat
T_{\kappa s}\vac= 0
\label{eq:Tanhilates}
\ee
so that finally
\be 
[\hat A, \hat R_{\kappa s}]=a_\kappa\, \hat R_{\kappa s} + \hat T_{\kappa s}
\label{eq:newcommuteA} 
\ee
with the $\hat T_{\kappa s}$ operators anhilating the vacuum.  The appearance
of these new operators requires one to understand something about their
commutation relations.

Since under rotations and gauge transformations $\hat T_{\kappa s}$ transforms
just like $\hat R_{\kappa s}$, one can take the commutators of $\hat T_{\kappa
s}$ with $\hat J_z, \hat J_\pm,$ and $\hat Q$ to parallel 
Eqs.~\ref{eq:commuteJz}-\ref{eq:commuteJ+} and \ref{eq:charge}. 
Then by the same argument that led to Eqs.~\ref{eq:three}, one finds that
$[\hat A, T_{\kappa s}]$ transforms just like $R_{\kappa s}$.   Now since by
Eq.~\ref{eq:newcommuteA} $[\hat A, R_{\kappa s}]$ can be replaced by $\hat
T_{\kappa s}$ and $\hat R_{\kappa s}$, then by Axiom 3 $[\hat A, T_{\kappa
s}]$ must be expressible as a linear combination of the operators
$\hat A,\ \hat {\bf J},\ \hat Q,\ \hat R_{\kappa s} $ and $\hat T_{\kappa s}$
which transforms like $[\hat A, T_{\kappa s}]$ under rotations and gauge
transformations.  The generic one is
\be
[\hat A, \hat T_{\kappa s}]\,= \sum_{n_\lambda\, t} \left( B_{\kappa
s}{}^{\lambda t}\ \hat T_{\lambda t} +  C_{\kappa s}{}^{\lambda
 t}\ \hat R_{\lambda t}\right)\\
+a_\kappa\,\delta_q{}^0\left[\delta_j{}^0\, (D\hat Q +
 E\hat A) + \delta_j{}^1\, F\hat J_{m}\right]
\label{eq:commuteT}
\ee
Here $\kappa$ and $\lambda$ share common $j, m$ and $q$, the coefficients
$B_{\kappa s}{}^{\lambda t}, C_{\kappa s}{}^{\lambda t}, D, E$ and $F$ are
structure constants ($D$ and $E$ may depend on $n_\kappa$, $s$ and
$F$ on $m_\kappa$ as well), and
$\hat J_m$ are the spherical tensor components of the {\it vector\ operator\/}
$\hat {\bf J}$, namely $\hat J_{\pm}/\surd 2$ and
$\hat J_z$.  The prefactor $a_\kappa$ is added for later convenience.

Upon operation with Eq.~\ref{eq:commuteT} on $\vac$, the only surviving terms
are those on the r.h.s. involving the $R_{\kappa s}$ because of
Eq.~\ref{eq:Tanhilates} and the fact that the vacuum bears no charge or angular
momentum and has zero area eigenvalue.  Thus necessarily $C_{\kappa
s}{}^{\lambda t}\equiv 0$ because $\hat R_{\lambda t}\vac$ cannot vanish.

One can also give an informal argument that either all $\hat T_{\kappa
s}\equiv 0$ or all $B_{\kappa s}{}^{\lambda t}\equiv 0$.  Consider a pair of
orthogonal  one--black hole states, $\ket{\negthinspace x\ }$ and
$\ket{\negthinspace y\ }$, of the sort $\hat R_{\kappa s}\vac$.  The matrix
element of Eq.~\ref{eq:commuteT} between these two states is
\be
\sum_{n_\lambda t} B_{\kappa s}{}^{\lambda t}\ \bra{\ y}\hat T_{\lambda
t}\ket{\negthinspace x\ } = (a_y-a_x)\bra{\ y}\hat T_{\kappa
s}\ket{\negthinspace x\ }
\label{eq:element1}
\ee
because $\bra{\ y}\hat Q\ket{x\ }$, {\it etc.} drop out by the orthogonality
of $\ket{\negthinspace x\ }$ and $\ket{\negthinspace y\ }$.  It is necessary
to adjust $q_y=q_x+q_\lambda$ so that $\bra{\ y}\hat T_{\lambda
t}\ket{\negthinspace x\ }$ shall not vanish trivially.  This follows from the
fact that in analogy with Eq.~\ref{eq:commuteQ}, $[\hat Q, \hat T_{\lambda t}]
= q_\lambda e\hat T_{\lambda t} $.  The matrix element of this last equation is
$(q_y-q_x-q_\lambda)\bra{\ y}\hat T_{\kappa s}\ket{\negthinspace x\ } = 0$, which
upholds the claim.  In like way one must take $m_y=m_x+m_\lambda$.

According to Eq.~\ref{eq:element1}, the ``vector'' whose components are
$\bra{\ y}\hat T_{\lambda t}\ket{\negthinspace x\ }$ for all $\lambda, t$ is
an eigenvector of the matrix $B_{\kappa s}{}^{\lambda t}$.  This is true for
every $\ket{\negthinspace x\ }$ if one adjusts $\ket{\negthinspace y\ }$ in
accordance with the mentioned constraints.  But in a framework where we
truncate the infinite dimensional problem to a finite dimensional one, the
number of such eigenvectors exceeds the dimension of $B_{\kappa s}{}^{\lambda
t}$ since for every pair $\kappa, s$ one can choose a $\ket{\negthinspace x\
}$ with the same quantum numbers, but then one is still free to choose $a_y$
and
$s_y$ in specifying $\ket{\negthinspace y\ }$.  The surplus may mean that the
eigenvectors constructed as above often vanish.  For instance, $\bra{\ y}\hat
T_{\kappa s}\ket{\negthinspace x\ }$ might vanish unless $a_y=a_x$ and
$s_y=s_x$.  Then we would have exactly the right number of (nontrivial)
eigenvectors.  But according to Eq.~\ref{eq:element1}, all the corresponding
eigenvalues would then vanish.  A matrix all whose eigenvalues vanish must
vanish, and so in this eventuality 
$B_{\kappa s}{}^{\lambda t}=0$.

One can escape the above conclusion if the $\bra{\ y}\hat T_{\kappa
s}\ket{\negthinspace x\ }$ with $\ket{\negthinspace y\ }$ properly adjusted
as above is the same vector up to normalization for several
different $\ket{\negthinspace x\ }$'s.  This does not look likely.  The
other escape clause is for all the  $\bra{\ y}\hat T_{\kappa
s}\ket{\negthinspace x\ }$ to vanish, which by completeness of the states
$\ket{\negthinspace x\ }$ and $\ket{\negthinspace y\ }$ means that $\hat
T_{\kappa s}=0$.  Then Eq.~\ref{eq:element1} is satisfied trivially.  In both
of the eventualities, the term involving $B_{\kappa s}{}^{\lambda t}\ \hat
T_{\lambda t}$ in Eq.~\ref{eq:commuteT} drops out. 

Accepting this one defines a new creation operator
\be
\hat R_{\kappa\, s}^{\rm new}\equiv  \hat
R_{\kappa\, s} + 
(a_\kappa)^{-1}\left\{\hat
T_{\kappa\, s} + \delta_{q_\kappa}{}^0\big[\delta_{j_\kappa}{}^0\, (D\hat Q +
E\hat A) + \delta_{j_\kappa}{}^1\, F\hat J_m\big]\right\}
\label{eq:newR}
\ee
Since $\hat T_{\kappa\,s}, \hat A, \hat J_m$ and $\hat Q$ all anhilate
$\vac$, it is seen that $\hat R_{\kappa\, s}^{\rm new}$ creates the same
one--black hole state as $\hat R_{\kappa\, s}$.  But the $\hat R_{\kappa\,
s}^{\rm new}$ turn out to satisfy simpler commutation relations.  Substituting
in $[\hat A, \hat R_{\kappa\, s}^{\rm new}]$ from Eqs.~\ref{eq:newcommuteA}
and \ref{eq:commuteT} one gets the commutator
\be 
[\hat A, \hat R_{\kappa\, s}^{\rm new}]=a_\kappa\, \hat R_{\kappa\, s}^{\rm
new}
\label{eq:newnewcommuteA} 
\ee
which is  reminiscent of Eqs.~\ref{eq:commuteJz} and \ref{eq:commuteQ}. 
Henceforth I use only $\hat R_{\kappa\, s}^{\rm new}$ but drop the ``new''.

\section{Algebraic Derivation of the Area Spectrum}\label{sec:last} 

Operating with $R_{\kappa s}\hat R_{\lambda t}$ on $\vac$
and simplifying the result with Eq.~\ref{eq:newnewcommuteA} gives
\be 
{ \hat A\hat R_{\kappa s}\hat R_{\lambda t}\vac =\hat R_{\kappa s}(\hat
A+a_\kappa)\hat R_{\lambda t}\vac =(a_\kappa+a_{\lambda}) \hat R_{\kappa
s}\hat R_{\lambda t}\vac }
\label{eq:commute} 
\ee
so that the state $\hat R_{\kappa s}\hat R_{\lambda t}\vac$ has horizon area
equal to the  sum of the areas of the states $\hat R_{\kappa s}\vac$ and $\hat
R_{\lambda t}\vac$.  Analogy with field theory might lead one to believe that
$\hat R_{\kappa s}\hat R_{\lambda t}\vac$ is just a two--black
hole state, in which case the result just obtained would be trivial. 
But in fact, the axiomatic approach allows other possibilities.

Recall Eqs.~\ref{eq:commuteJz}, \ref{eq:commuteQ} and
\ref{eq:newnewcommuteA}, namely 
\be
[\hat X, \hat R_\kappa] = x_\kappa \hat R_\kappa\qquad {\rm for}\qquad \hat X
=\{\hat A, \hat Q, \hat J_z\}
\label{eq:commutators}
\ee
The Jacobi identity, Eq.~\ref{eq:Jacobi}, can then be used to infer that
\be
[\hat X, [\hat R_{\kappa}, \hat R_{\lambda}]] = (x_\kappa+x_\lambda) [\hat
R_{\kappa},\hat R_{\lambda}]
\label{eq:commutesummary}
\ee
which makes it clear that $[\hat R_{\kappa}, \hat R_{\lambda}]$ has the same
transformations under rotations and gauge transformations as a single $\hat
R_{\mu}$ with
\be
 x_\mu\equiv x_\kappa+x_\lambda 
\label{eq:additivity}
\ee 
Axiom 3 then allows one to conclude that ($\varepsilon_{\kappa\lambda}$ and
$e_{\kappa\lambda}$ are structure constants)
\be
[\hat R_{\kappa}, \hat R_{\lambda}]
 =\sum_{\mu} \left(\varepsilon_{\kappa\lambda}{}^{\mu}\, \hat
R_{\mu} + e_{\kappa\lambda}{}^{\mu}\, \hat T_{\mu}\right)
+ a_\mu\delta_q{}^0\left[\delta_j{}^0\,
(\tilde D\hat Q + \tilde E\hat A) + \delta_j{}^1\, \tilde F\hat J_m\right]
\label{eq:twoR}
\ee
where $j, m, q\in \mu$. Although closure was postulated with respect to the
old $\hat R$'s, we use the new $\hat R$'s here.  This causes no difficulty
because the two differ only by a superposition of $\hat T$'s, and such terms
have been added anyway.

When one operates with Eq.~\ref{eq:twoR} on $\vac$ one gets
\be
[\hat R_{\kappa}, \hat R_{\lambda}]\vac\, = \,
\ket{\negthinspace\bullet\ }
\label{eq:one}
\ee
where $\ket{\negthinspace\bullet\ }$ stands for a {\it one\/}--black hole
state, a superposition of states with various $\mu\, t$.  Were
$\hat R_{\kappa s}\hat R_{\lambda t}\vac$ purely a two--black hole state
as suggested by the field--theoretic analogy, one could not get
Eq.~\ref{eq:one}.  Inevitably
\be
\hat R_{\kappa s}\hat R_{\lambda
t}\vac\,=\,\ket{\negthinspace\negthinspace\bullet\bullet\ }\, +
\,\ket{\negthinspace\bullet\ }
\label{eq:two}
\ee
with $\ket{\negthinspace\negthinspace\bullet\bullet\ }$ a two--black hole
state, symmetric under exchange of the $\kappa s$ and $\lambda t$ pairs. 
The superposition of one and two--black hole states means that the rule of
additivity of eigenvalues, Eq.~\ref{eq:additivity}, applies to one black hole
as well as two: {\it the sum of two eigenvalues of $\hat Q$,  $\hat J_z$ or
$\hat A$ of a single black hole is also a possible eigenvalue of a single black
hole.\/}   For charge or $z$--spin component this rule is consistent with
experience with quantum systems whose charges are always integer multiples of
the fundamental charge (which might be a third of the electron's), and whose 
$z-$ spins are integer or half integer multiples of $\hbar$.  This agreement
serves as a partial check of our line of reasoning.

In accordance with Axiom 1, let $a_1$ be the smallest nonvanishing eigenvalue
of $\hat A$. Then Eq.~\ref{eq:additivity} says that any positive
integral multiple $na_1$ (which can be obtained by repeatedly adding $a_1$ to
itself) is also an eigenvalue.  This spectrum of $\hat A$ agrees with that
found in Sec.~\ref{sec:demystify} by heuristic arguments.  But the question
is, are there any other area eigenvalues in between the integral ones (this
has a bearing on the question of whether splitting of the levels found in
Sec.~\ref{sec:demystify} is at all possible) ?

To answer this query, I write down the hermitian conjugate of
Eq.~\ref{eq:newnewcommuteA}:
\be
[\hat A, \hat R_{\kappa}^\dagger]=-a_\kappa \hat R_{\kappa}^\dagger
\label{eq:conjugate}
\ee
Then
\be
\hat A\, \hat R_{\kappa}^\dagger \hat R_{\lambda}\,
\vac =\, \left(\hat R_{\kappa}^\dagger\hat A - a_\kappa
\hat R_{\kappa}^\dagger
\right)\hat R_{\lambda} \vac\, =
\left(a_{\lambda}-a_\kappa\right) \hat R_{\kappa}^\dagger 
\hat R_{\lambda}\vac
\label{eq:commuteconj}
\ee
Thus differences of area eigenvalues appear as eigenvalues in their own
right.  Since $\hat A$ has no negative eigevalues, if $n_\lambda\,\leq
\,n_\kappa$, the operator $\hat R_{\kappa}^\dagger$ must anhilate the
one--black hole state $\hat R_{\lambda}\, \vac$ and there is no black
hole state  $\hat R_{\kappa}^\dagger \hat R_{\lambda}\, \vac$.  By contrast,
if $n_\kappa\,<\,n_\lambda$, $R_{\kappa}^\dagger$ obviously lowers the area
eigenvalue of $\hat R_{\lambda}$.  There is thus no doubt that $\hat
R_{\kappa}^\dagger \hat R_{\lambda}\, \vac$ is a purely one--black hole state
(a ``lowering'' operator cannot create an extra black hole:
Eq.~\ref{eq:commuteconj} shows that $\hat R_{\kappa}^\dagger$ anhilates the
vacuum). In conclusion, {\it positive differences of one--black hole area
eigenvalues are also allowed area eigenvalues for one black hole.\/}  

If there were fractional eigenvalues of $\hat A$, one could, by substracting a
suitable integral eigenvalue, get a positive eigenvalue below $a_1$, in
contradiction with $a_1$'s definition as lowest positive area eigenvalue. 
Thus the set $\{na_1;\ n=1, 2, \cdots\}$ comprises the totality of
$\hat A$ eigenvalues for one black hole, in complete agreement with the
heuristic arguments of Sec.~\ref{sec:demystify} (but the algebra by itself
cannot set the area scale $a_1$).

What about the degeneracy of area eigenvalues ? According to Axiom 1, $g(n)$,
the degeneracy of the area eigenvalue $na_1$, is independent of $j, m$ and
$q$.  Thus for fixed $\{n_\kappa, j_\kappa, m_\kappa, q_\kappa\}$ where not
all of $j_\kappa, m_\kappa$ and $q_\kappa$ vanish, there are
$g(n_\kappa)$ independent one--black hole states $\hat R_{\kappa\,s}\vac$
distinguished by the values of $s$.  Analogously, the set $\{n_\lambda=1,
j_\lambda=0, m_\lambda=0, q_\lambda=0\}$ specifies $g(1)$ independent
states $\hat R_{\lambda\,t}\vac$, all different from the previous ones
because not all quantum numbers agree.  One can thus form
$g(1)\cdot g(n_\kappa)$ one--black hole states, $[\hat R_{\kappa\,s},
\hat R_{\lambda\,t}]\vac$, with area eigenvalues $(n_\kappa + 1)a_1$ and
charge and angular momentum just like the states $\hat R_{\kappa\,s}\vac$. 
{\it If\/} these new states are independent, their number cannot exceed the
total number of states with area $(n_\kappa + 1)a_1$, namely $g(n_\kappa +1)
\geq g(1)\cdot g(n_\kappa)$.  Iterating this inequality
starting from
$n_\kappa=1$ one gets
\be
g(n) \geq  g(1)^n
\label{eq:degeneracy2}
\ee
The value $g(1)=1$ is excluded because one knows that there
is some degeneracy.  Thus the result here is consistent with the law
\ref{eq:degeneracy} which we obtained heuristically.  In particular, it 
supports the idea that the degeneracy grows exponentially with area.  The
specific value
$g(1)=2$ used in Sec.~\ref{sec:demystify}  requires further input.

\section*{Acknowledgments}

I thank Slava Mukhanov for inspiring conversations and Avraham Mayo for
discussions.  This research is supported by a grant from the Israel Science
Foundation established by the Israel National Academy of Sciences.

\end{document}